# Computational Grids in Action: The National Fusion Collaboratory

K. Keahey[1]  T. Fredian[2]  Q. Peng[3]  D. P. Schissel[3]  M. Thompson[4]  I. Foster[1,5]  M. Greenwald[2]  D. McCune[6]

[1] Mathematics and Computer Science Division, Argonne National Laboratory, Argonne, IL 60439
[2] Plasma Science and Fusion Center, MIT, Cambridge, MA 02139
[3] General Atomics, P.O. Box 85608, San Diego, California, 92186-5608
[4] Lawrence Berkeley National Laboratory, Berkeley, CA 94720
[5] Department of Computer Science, University of Chicago, Chicago, IL 60657
[6] Princeton Plasma Physics Laboratory, Princeton, NJ 08543-0451

## 1. Introduction

The term computational Grids [1] refers to infrastructures aimed at allowing users to access and/or aggregate potentially large numbers of powerful and sophisticated resources. More formally, Grids are defined as infrastructure allowing flexible, secure, and coordinated resource sharing among dynamic collections of individuals, institutions and resources referred to as *virtual organizations* [2]. In such environments, we encounter unique authentication, authorization, resource access, and resource discovery challenges. Each new Grid application facilitates a deeper understanding of these challenges and takes the Grid technology to increasing levels of functionality, performance, and robustness. In this paper we describe a new Grid application: the recently funded U.S. National Fusion Collaboratory and its potential for driving the understanding and development of computational Grids.

The National Fusion Collaboratory project [3] was created to advance scientific understanding and innovation in magnetic fusion research by enabling more efficient use of existing experimental facilities through more effective integration of experiment, theory, and modeling. Magnetic fusion experiments operate in a pulsed mode producing plasmas of up to 10 seconds duration every 10 to 20 minutes, with multiple pulses per experiment. Decisions for changes to the next plasma pulse are made by analyzing measurements from the previous plasma pulse (hundreds of megabytes of data) within roughly 15 minutes between pulses. Clearly, this mode of operation could be made more efficient by the ability to do more analysis in a short time. Transparent access to remote high-powered remote computational resources, data, and analysis codes would allow more researchers to perform more analysis thus enabling more effective comparison of theory and experiment both during experiments and afterwards. Furthermore, such access would encourage sharing of software and hardware resources allowing the researchers to provide them as network services to the community. In this paradigm, access to resources (data, codes, visualization tools) would be separated from their implementation, freeing the researcher from needing to know about software implementation details and allowing a sharper focus on the physics.

These requirements make the Fusion Collaboratory an interesting example of a virtual organization. To realize its goals the Collaboratory aims to create and deploy collaborative software tools throughout the national magnetic fusion research community, which comprises over one thousand researchers from over forty institutions (Figure 1). In building a Fusion Grid, the Collaboratory will work with the Globus Project™[1] on deploying and extending its Globus Toolkit™ [4], and applying the Akenti [5] authorization policies in the Grid context. New technology developments are also required; in particular to provide the quality of service guarantees needed for the near real-time execution of data analysis and simulation jobs during fusion experiments. To this end, project participants are investigating advance reservations of multiple resources (such as computational cycles, possibly network bandwidth), the implementation of priority-based policies for jobs, and job pre-emption in collaboration with local schedulers.

---
[1] Globus Project and Globus Toolkit are trademarks held by the University of Chicago.

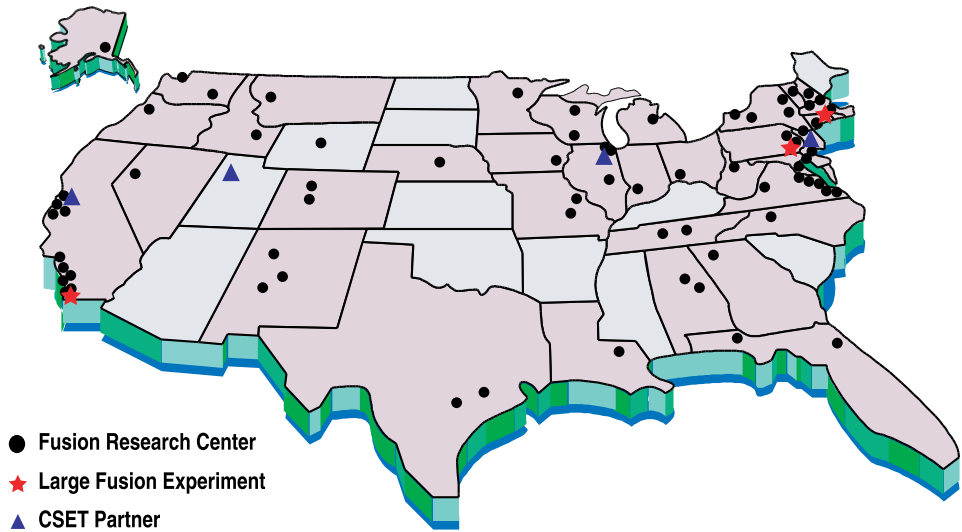

- **Fusion Research Center**
- ★ **Large Fusion Experiment**
- ▲ **CSET Partner**

**Figure 1. Magnetic Fusion Research is conducted throughout the United States.**

This article describes in detail the needs of the fusion community, the infrastructure enabling virtual organizations provided by the Globus Toolkit, and the Akenti project's authorization and use policies which will be used to fulfill resource management goals of the collaboratory. We then discusses the problems we are trying to solve, outline some proposed solutions, and describe preliminary results obtained in the first stage of our work, in which we used the Globus Toolkit to Grid-enable a fusion application.

The structure of the paper is as follows. Section 2 describes the Fusion experiments, their mode of operation, and the application-specific software. Section 3 reviews the Grid technologies that we are using to create the collaboratory. Sections 4 and 5 present preliminary results of our work and describe how the Grid technologies will be used to address the challenges of the collaboratory. In section 6 we briefly summarize the visualization research carried out as part of this project, and conclude in section 7.

## 2. Magnetic Fusion Applications and Frameworks

Magnetic fusion research is concerned with the design and construction of fusion reactors, with the ultimate goal of creating commercially viable systems. In pursuit of this goal, the fusion community not only conducts research on alternative reactor designs but also performs experiments centered at three large facilities: Alcator C-Mod [6], DIII-D [7], NSTX [8]. Data produced by both simulations and experiments are an important community resource. As described earlier, magnetic fusion experiments operate in a pulsed mode, with many pulses a day produced at roughly 15 minute intervals. For each pulse up to 10,000 separate measurements versus time are acquired representing over 250 MB of data. Decisions for changes to the next plasma pulse are based on analyzing this data within the roughly 15 minute inter-pulse intervals. For this reason, the ability to produce and analyze large amounts of data between-pulses by a geographically dispersed research team has been recognized by the scientific community as leading to more efficient utilization of experimental facilities.

Teaming with this experimental community is a theoretical and simulation community that concentrates on the creation of realistic non-linear 3D plasma models. Although the fundamental laws that determine the behavior of fusion plasmas are well known, obtaining their solution under realistic conditions is a scientific problem of enormous complexity, due in large part to the enormous range of temporal and spatial scales involved. Such a challenge can be met only with advanced scientific computing. Because of this complexity, researchers have a long history of productive use of advanced computation and modeling. Working together to advance scientific understanding and innovation, these two groups represent over one thousand scientists from over forty institutions.

As a first step towards addressing the problems of data sharing and rapid data analysis the fusion community adopted a common data acquisition and management system, MDSplus [9] and a common relational database run-management schema, currently implemented in a Microsoft SQL Server. MDSplus was designed primarily to serve as a high-speed, hierarchical data acquisition system for fusion experiments. In addition to capabilities allowing it to accommodate both simple and complex data types, it also contains functionality enabling scientists to describe data acquisition and analysis tasks that need to be accomplished during an experimental pulse cycle, and dispatching these tasks to distributed servers. For example, a user can establish complex sequencing of such tasks by describing dependencies between them (e.g., execute task C only if tasks A and B have completed successfully). MDSplus also provides a mechanism for defining events that should be declared when a particular task completes. Applications can notice the occurrence of these events and respond with some operation such as refresh a graphical display using the newest data.

Although MDSplus data servers provide efficient mechanisms for data storage and retrieval, they are difficult and inefficient to search by content. For this reason, their functionality was supplemented by calls to the relational database providing a fast search capability. The relational database stores metadata information about the data produced by a program as well as information on how to find that data in the MDSplus server. The metadata contains information about who produced the data, what the data represents, and how and when it has been produced, and can optionally be augmented by highlights of the data. To retrieve a specific piece of data the scientist first searches the relational database, and having found information about the data he or she needs, can take advantage of the fast retrieval option provided by MDSplus data server associated with this database. This storage system has proved especially valuable in organizing numerous data repositories generated by complex theoretical simulations. It is expected that in the future the data will be stored in several different MDSplus servers, in which case the metadata will be extended to include the address of a server.

Over the years, the fusion community has developed many data analysis and theoretical simulation codes shared by a geographically diverse users group. There is an ongoing effort to convert these codes to use a common data format supported by the MDSplus. In a typical scenario, a code would identify the data it requires through the relational database, read it from the associated MDSplus server, perform its computations, and then update both the metadata information and the data on the MDSplus server. The action of writing the data will cause the generation of an MDSplus event signaling the availability of the data. The data can then be read by another code for further analysis or visualization. In the context of this interaction, an important further advantage of using MDSplus is that it provides a historical archival capability and therefore solves the problem of accidentally overwriting, losing, or destroying data.

Although this infrastructure significantly advances the ability to process data produced by fusion experiments, more capabilities are required in a truly distributed environment in which several laboratories come together to work as one virtual organization. In the context of such a collaboratory, security, authorization, and use policies applying to both data and resources acquire critical importance, as the levels of trust between institutions need to be more formally defined and enforced. Further, near real-time execution and result delivery capabilities required by the nature of fusion experiments need to be guaranteed. Also, separation of data and metadata, already embraced by the fusion community, can be exploited further by leveraging mechanisms provided by Data Grids [10] in order to improve efficiency of interactions. Finally, introducing these new capabilities into the collaboratory will allow the fusion scientists to offer the fusion codes, together with hardware, software and maintenance involved as a service to the community. Such services, combined with user-friendly mechanisms for accessing them will allow the scientist to focus on physics and lower the entry barrier into the collaboratory.

3. **Grid Technologies**

The Globus Project has been working for several years to develop capabilities enabling flexible, secure, and coordinated resource sharing between virtual organizations. The result of these efforts, the Globus Toolkit, is an open-architecture, open-source set of protocols, services and tools that address central problems in Grid computing, including security, resource discovery, remote resource access, and data access.

The *Grid Security Infrastructure* (GSI) [11] provides for "single sign-on" authentication, delegation of user credentials to application programs, and local control over authorization. GSI uses X.509 identity certificates [12] passed over Transport Layer Security (TLS) [13] for secure authentication. GSI credentials can also be used to establish secure (authenticated, integrity-checked and/or encrypted) communication channels. Single-sign-on authentication means that once a user has logged into the Grid, subsequent security measures within the Grid are transparent; in other words by delegating credentials programs can be run on the user's behalf without the need to separately authenticate every action.

The *Meta Directory Service* (MDS) [14] provides GSI-enabled access to information pertaining to the availability and state of Grid resources. MDS provides an information infrastructure for computational Grids using the Lightweight Directory Access Protocol (LDAP) as a uniform means of querying system information from a rich variety of system components, and for constructing a uniform namespace for resource information across multiple organizations.

The *Globus Resource Allocation and Management* GRAM [15] service defines a protocol (and API) for secure remote submission, monitoring, and management of compute jobs. GRAM provides a uniform interface to various local resource management tools such as PBS, LSF, NQE, LoadLeveler, and Condor. It uses a flexible Resource Specification Language (RSL) to provide a method for exchanging information about resource requirements, job process creation, and job control between all of the components in the GRAM architecture. RSL requests are communicated to a remote resource where they can be either denied (based on evaluating a use policy for example) or accepted. In the latter case, GRAM also enables remote monitoring (providing for example estimated delay times) and management of the created jobs.

Another set of Globus Toolkit "Data Grid" technologies support the distributed management of large amounts of data. GridFTP [16] provides an efficient, high-performance, reliable, secure, and policy-aware implementation of large-scale data movement. Replica catalog and replica management tools [17] deal with issues of replicating data and locating replicas based on the metadata information using a strategy similar to, but much more performance-oriented than, that already present in MDSplus.

Although the Globus Toolkit already provides much of the functionality needed by the fusion community, important capabilities are missing. In particular, the near-real-time requirements of fusion are motivating extensions to the GRAM protocol—and to scheduler implementations—to support advanced reservations, pre-emption, and quality of service.

Another area where extensions are required relates to local authorization. In the current Globus Toolkit, a GRAM server checks authorization by consulting a local access control list. Yet a simple access control list is not a good means of maintaining the access policies for a virtual organization such as the fusion collaboratory. It is important that these access policies be easily managed by the resource stakeholders, some of whom may be research scientists located remotely from the resource, for example, in the case of shared data. The Akenti authorization service emphasizes convenient specification of access policies for resources by multiple remote stakeholders. Akenti enforces access policy that is contained in certificates that are digitally signed by the private keys of users identified by X.509 certificates. Thus access policy can be created and stored remotely from a resource, since a signed document can be verified at use time to ensure that it was signed by the issuer, and that it has not been modified since it was signed. Akenti policy can allow access to any arbitrary hierarchical or flat sets of resources. Thus, the access policy for both compute cycles and data can be set in a uniform way, using a single user identity for each collaboratory member. It can allow access based on components of a user's Distinguished Name (as contained in his identity certificate), or by groups, roles or any other stakeholder defined category, and supports additional runtime constraints such as time-of-day, level of machine usage or IP address from which the request originated. We propose to apply, and if necessary extend, Akenti in the fusion collaboratory, as well as investigating the utility of the related Community Authorization Service.

## 4. Preliminary Experiments and Results

The interest of the fusion community in the use of computational Grids is demonstrated by the fact that very shortly after its inception the Collaboratory was able to implement a prototype of a typical fusion application running in the Grid environment. More specifically, we wished to demonstrate that we could allow a fusion scientist to sign on at their desktop and then run a simulation code and access MDSplus data at two distinct remote locations. Although of limited capability, this application is significant as a first step towards a capability that fusion scientists expect the Grids to facilitate. This prototype was demonstrated at the SC'2001 conference in November 2001.

The prototype is composed of four components interacting through a Globus-enabled implementation of the MDSplus framework:
- A controller program orchestrating the interactions of the components, and interacting with the user through a Graphical User Interface (GUI)
- An MDSplus data server where data is stored and retrieved from
- EFIT [18] , an MPI-based parallel analysis program for magnetic equilibrium reconstruction

- Scientific visualization of the data

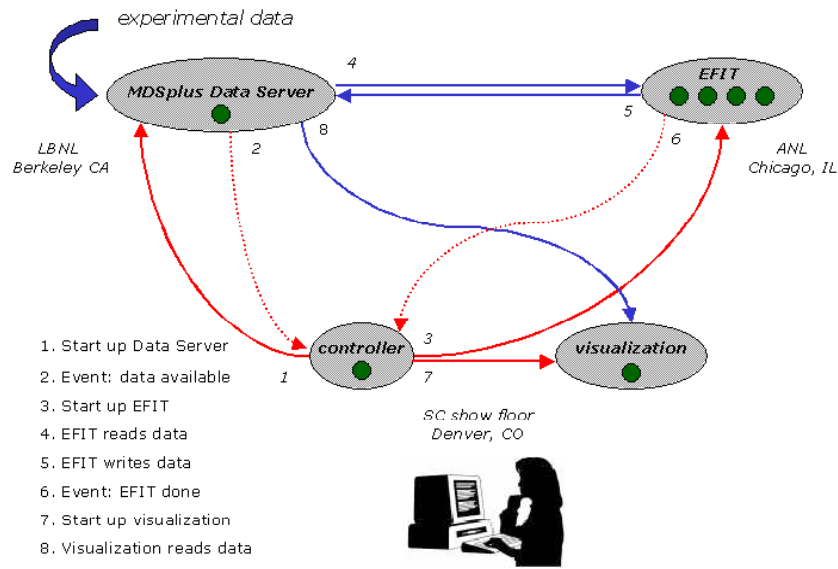

**Figure 2. Interactions in SC2001 Fusion Collaboratory Demonstration**

The interactions are depicted in Figure 2. After logging into the Fusion Grid using Globus Toolkit credentials, the fusion scientist starts the controller program on a workstation. Using the scientist's credential the controller then activates a data server on a remote machine and waits for the experimental data to become available. When this happens, the controller is notified by an MDSplus event generated by the data server. Proceeding on this notification, the controller remotely starts up the parallel EFIT analysis code using the Globus GRAM protocol. After the analysis is completed, EFIT writes the data to the data server and sends a completion event to the controller. When the controller sees the completion event, it starts up the visualization program on the scientist's workstation, which again reads data from the data server, and visualizes it. During the SC 2001 demonstration the controller and visualization components were running at the demonstration workstation on the SC show floor in Denver, the data server controller was running at Lawrence Berkeley National Laboratory (LBNL) while the EFIT analysis code was running at Argonne National Laboratory (ANL).

This application was Grid-enabled by replacing MDSplus socket calls by calls to the Globus IO library that offer authentication and secure data transfer under familiar socket API. In addition, the Globus GRAM tool was used by the controller to start, monitor, and manage remote computations. Overall, the amount of work necessary to Grid-enable the application was relatively small. The necessary work was carried out in the space of roughly a month, primarily by fusion scientists expert in application-specific software but with no previous exposure to the Globus Toolkit. The rapid development of this prototype was largely enabled by the fact that we managed to leverage rather than replace the existing MDSplus software underlying the participating fusion applications. It was also facilitated by the flexible nature of the Globus Toolkit™ which allows applications to be Grid-enabled incrementally; in this instance for example we added the GSI-based interactions first, and later automated the resource management process by using GRAM.

From the point of view of the Fusion Collaboratory, the main capabilities that we demonstrate here are multi-site security and uniform access to remote computational capabilities. Adding these capabilities means that this interactive scenario can now leave the domain of one fusion laboratory and instead be executed across multiple fusion sites in a truly collaborative fashion. It is also worth noting that instrumenting the MDSplus with calls to Globus IO lead to the development of general-purpose functionality that will be used to Grid-enable other fusion applications.

## 5. Building the Fusion Grid

We envision our work on the fusion collaboratory as a progression of testbeds, each building on and extending the capabilities available in the previous one. In the previous section we described our preliminary experiments with building Grid-enabled Collaboratory infrastructure. These experiments provided a proof of concept for our design of integrating Globus with the existing fusion frameworks. Building on these results we will build the Fusion Grid in the following phases.

### 5.1 Deployment Phase

Our next step will be to deploy Globus on fusion sites and adapt it to their needs. Initially the Fusion Grid will be composed from resources provided by the three Fusion Laboratories: General Atomics (GA), Princeton Plasma Physics Laboratory (PPPL), and the Massachusetts Institute of Technology (MIT) Plasma Science and Fusion Center. The main compute engine of the Fusion Grid is a 60 node Beowulf Linux cluster contributed by PPPL, and each of the laboratories also contributes a database server. Additional computational resources, forming part of the DOE Science Grid, can be contributed later as need arises by Argonne National Laboratory (ANL), Lawrence Berkeley National Laboratory (LBNL), and perhaps other DOE laboratories.

Given that the basic Grid infrastructure is already in place, our first step towards realizing the Collaboratory's goals is to build a Fusion Grid. This step involves the deployment of Globus on participating facilities and the implementation of interoperability solutions to support GSI-enabled access to resources that do not already support GSI. We obtain X.509 identity certificates credentials for each collaboratory user and server from a Certificate Authority operated by ESnet for the DOE Science Grid. Credentials issued by this CA can be used solely within the context of a single collaboratory or alternatively may also be used to gain access to other virtual organizations of which an individual may be a member. In this phase we will also resolve the affinities between the existing MDSplus software and Globus identified in our initial work, so as to best leverage the features of both.

As a result of this step many of the Collaboratory's goals will be fulfilled: the scientists will be able to securely run codes on remote resources within the Fusion Grid, and access fusion data in databases throughout sites belonging to the Collaboratory. In other words, at this stage the fusion scientists should be able to start using the Grid for at least some of their research.

### 5.2 Use Policies and Issues of Trust

In the next phase of our work will address the issue of trust within the collaboratory. For example, some experimental data are meant to be shared with only a subset of the community or through managed levels of access. Similarly, sharing resources also needs to be restricted to provide fair use and prioritize it with respect to community goals. In this phase we will work with Akenti to integrate authorization and use policies into Globus, and will develop the Globus monitoring and accounting capabilities to fulfill the Collaboratory's needs.

The fusion community already allows data sharing though the MDSplus data servers, but access control methods that are used require individual user accounts and passwords for each of the resources that are needed. The access control policies are all set local to the resource, in resource specific ways. GSI generalizes and extends this process through assigning to each user a single credential that is uniformly used in all authorization decisions and conducting all communication between users though authenticated communication channels. Adding access control based on Akenti policies will provide further extensions. These policies for resource use can be viewed by remote stakeholders and other authorized users. Note that fine-grained access control may still be accomplished by mapping a Grid identity onto a local user account.

To begin with, we will experiment with simple policies either granting a user access to a resource or denying it. Later, we can extend Akenti to express more complex user policies in terms of for example CPU cycles that individual users can be granted (number of nodes, individual or cumulative run-time), or storage grant on a given resource. These policies will also cover coping strategies, which can later be used to ensure for example priority and preemption. Similar policies will be applied to MDSplus data retrieval. A complementary thrust in this phase of the project is to monitor and audit resource use to make sure that the policies are indeed executed. This will be achieved through access logging and using capabilities that are either already present in Globus or will be developed in this phase.

The testbed resulting from these activities should enable fusion scientists to trust the Grid in the sense that their computations will indeed make progress as promised, and their data will not be compromised. We hope that this will open the Fusion Grid to more, and more critical applications.

**5.3  Moving to Real-Time**

Having put in place infrastructure whereby we can assign and enforce priority for certain jobs we will proceed to the next phase of the project, in which we focus on providing real-time guarantees for jobs associated with an ongoing experiment. Here, we will experiment with preemption issues for high-priority jobs as well as advance reservations for those runs.

Our work in the third phase of the project will build on authorization policies developed earlier. As an example, an access policy requirement that is unique to this collaboratory is the need to give high priority to any between-pulse computations. These jobs must be able to take precedence in use of compute resources, network bandwidth and data access over other less time-critical jobs. Such policy can then be enforced by Globus job execution and scheduling mechanisms implemented in GRAM with the collaboration of local schedulers (for example in order to implement job preemption where an existing job is suspended in favor of those of higher priority).

This work will result in a production Fusion Grid that can actually be used by fusion scientists to conduct data analysis during an ongoing experiment. In a larger context, it will strongly contribute towards the goal of enabling more real-time analysis in less time and thereby more efficient use of experimental facilities.

**5.4 Wrapping It Up**

Finally, having experienced the full range of interactions of fusion applications relevant to the scope of this work, we will design interfaces allowing the fusion scientists to provide resources and capabilities as services to the community.

Building a framework providing sophisticated resource management, monitoring, and run-time execution guarantees within the fusion community will provide a base on which even closer collaboration can be built. For example, it will enable communities to specialize in developing and maintaining codes dedicated to specific tasks, which would then be provided to the community as network services. This means that in order to run fusion codes the user will no longer have to go through the process of installing the software and all of its dependencies, applying patches etc., but will instead be able to run it directly through a convenient interface.

It is our hope that providing the infrastructure implementing this mode of operation will make the Fusion Grid more accessible to more fusion scientists which is the ultimate goal of this collaboratory.

**6.  Collaborative and Visualization Technologies**

Although the initial work described here concentrates on the creation of the Fusion Grid, a significant component of the Fusion Collaboratory involves scientific visualization. The demand placed on visualization tools by the Fusion Grid will be intense due to both the highly collaborative nature of fusion research and the dramatic increase in data resulting from the enhanced computational capabilities. One component of this work will be the creation of a collaborative control room for the three main experimental facilities. The collaborative control room concept allows the large on-site group to interactively work with small to large off-site groups. This work will focus on the use of high-resolution tiled display walls combined with an Access Grid node allowing a large group of scientists to explore information in collaboration more effectively than if they were crowded around a single workstation display. This technology will allow off-site research teams to be intimately involved in the decision making process during the important 15 minute between-pulse analysis cycle.

Control room collaboration involving between-pulse data analysis is the most time critical environment with fusion research and will therefore place the greatest stress on our emerging collaborative technologies during testing. Successful tests under control room conditions will mean that this technology will also work for less time critical applications such as working group meetings involving, for example, experimental data analysis, simulation code development, or the comparison of theory and experiment.

The other visualization component of the Fusion Collaboratory is the creation of new visualization tools for both experimental and simulation data that will represent a significant increase in capability and efficiency for the fusion community. Our goal is to provide users with visualizations that incorporate and compare data from experimental and simulation sources and to reflect uncertainty information to aid in data analysis and decision making. We seek to provide visualization capabilities that present a more complete and accurate rendition of data for users to analyze including novel methods for data uncertainty visualizations. Initial work on visualization was presented at SC 2001 where plasma temperature data was combined with magnetic field topology computed by EFIT to produce multiple time-dependent 3D electron temperature isosurfaces (Figure 3).

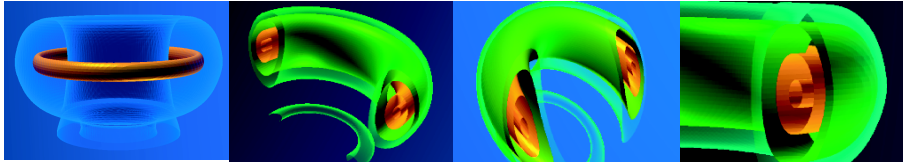

**Figure 3. Electron Temperature Isosurfaces (image courtesy of C. Johnson, University of Utah)**

## 7. Conclusions

The large scale and unique characteristics of the Fusion Collaboratory make it an important testbed for computational Grids. The fusion scientific community has well-defined needs for using Grids and has developed a software infrastructure that is a natural fit for Grid-based applications. In addition, the emphasis on near real-time constraints and the trend towards providing full-fledged services (rather than just hardware) to the community allow us to experiment with those features and develop Grid capabilities related to them.

Our first experiment in building a Fusion Grid is a proof of concept demonstrating that Globus Toolkit services can indeed be successfully used in this application. In a larger context, it shows that the Globus Project has developed a set of functionalities enabling virtual organizations. Further, the fact that significant functionality was added in very little time, that the development was carried out by application scientists, and that in the process we managed to leverage rather than replace existing application-specific infrastructure indicates that this functionality is provided in a compact and accessible way. Clearly, much credit is due to the fusion community whose recognition of the potential of distributed computing and acceptance of the distributed computing paradigm made it a perfect Grid-based application.

Although our experiment by no means represents the extent of work that still needs to be done to satisfy all of the goals of the Fusion Collaboratory, these preliminary results are very encouraging. A clear definition of challenging requirements will allow us to expand the capabilities of the Globus Toolkit in a truly application-driven way